\documentclass[11pt]{article} 
\usepackage[T1]{fontenc}
\usepackage[utf8]{inputenc} 

\usepackage{amsthm}
\newtheorem{theorem}{Theorem}

\usepackage{geometry} 
\usepackage{graphicx} 

\usepackage{booktabs} 
\usepackage{array} 
\usepackage{paralist} 
\usepackage{verbatim} 
\usepackage{subfig} 
\usepackage{amsmath} 
\usepackage{amssymb} 
\usepackage{url} 
\usepackage{epstopdf} 
\usepackage{color}
\usepackage{stackengine}
\usepackage{graphicx}
\usepackage{cite}
\usepackage{hyperref}

\usepackage{fancyhdr} 
\usepackage{sectsty}
\allsectionsfont{\sffamily\mdseries\upshape} 

\usepackage[nottoc,notlof,notlot]{tocbibind} 
\usepackage[titles,subfigure]{tocloft} 

\usepackage[T1]{fontenc}
\usepackage[utf8]{inputenc}

\title{Regularized  Stochastic Block Model for robust community detection in complex networks}
\author{Xiaoyan Lu$^{1}$, Boleslaw K. Szymanski$^{1,2}$\footnote{E-mail: szymab@rpi.edu}}

\date{}
\begin{document}
\maketitle

\begin{flushleft}
$^{\bf{1}}$ Social and Cognitive Networks Academic Research Center and Department of Computer Science, Rensselaer Polytechnic Institute, Troy NY 12180, USA\\
$^{\bf{2}}$ Spo\l{}eczna Akademia Nauk, \L{}\'{o}d\'{z}, Poland\\
\end{flushleft}

\section*{Abstract} 
The stochastic block model is able to generate different network partitions, ranging from traditional assortative communities to disassortative structures. Since the degree-corrected stochastic block model does not specify which mixing pattern is desired, the inference algorithms, which discover the most likely partition of the networks nodes, are likely to get trapped in the local optima of the log-likelihood. Here we introduce a new model constraining nodes' internal degrees ratios in the objective function to stabilize the inference of block models from the observed network data. Given the regularized model, the inference algorithms, such as Markov chain Monte Carlo, reliably finds assortative or disassortive structure as directed by the value of a single parameter. We show experimentally that the inference of our proposed model quickly converges to the desired assortative or disassortative partition while the inference of degree-corrected stochastic block model gets often trapped at the inferior local optimal partitions when the traditional assortative community structure is not strong in the observed networks.
\\
\newline
\textbf{Keywords:} community detection, stochastic block model, Markov chain Monte Carlo, modularity, assortative structures
\\

\section*{Introduction}
The study of modular structure in networks has a long history in the literature \cite{parnas1984modular,newman2004finding,newman2003structure,fortunato2010community}. The primary focus of this line of work has been the discovery of the assortative community structures where the edges inside each module are generally more numerous than the edges across them. In complex networks, the modular structures in networks include not only such assortative community structures but also other mixing patterns, like the core-periphery structures~\cite{borgatti2000models} and bi-partite structures. These various structures found in networks demand a more comprehensive class of community detection models than the assortative ones, targeted by the {\it modularity}-based approaches~\cite{newman2004finding}. Modularity maximization is perhaps the most widely used approach for community detection. It aims at maximizating the modularity of the network partitions, which is a broadly accepted quality metric comparing the number of edges observed in each community to the expected number in a random graph with the same degree sequence. As shown in~\cite{newman2016equivalence}, modularity maximization is, in fact, equivalent to the maximum likelihood estimation of the degree-corrected planted partition model, a special case of the degree-corrected stochastic block model~\cite{karrer2011stochastic}. Therefore, the stochastic block model can be considered more general than the traditional modularity-based community detection models because it can generate a wide variety of different network structures, which include not only the traditional assortative community structures, but also the disassortative structures such as the core-periphery structures~\cite{borgatti2000models}. Exceeding the limitations of modularity maximization approaches, the statistical inference of stochastic block model, which maximizes the likelihood of generating the observed network data, is capable of discovering of both assortative and disassortative structures in networks. The stochastic block model and its variants have gained significant attention in the past decade.

In the following, when we discuss the nodes' block assignment, for any node $l$, its block assignment is denoted by $g_l$. The standard stochastic block model is a variant of the generative random graph model which assumes the probability of connecting two nodes in a graph is defined exclusively by their block assignments. More precisely, this probability follows \textit{Bernoulli} distribution with the mean $\omega_{g_i,g_j}$. Hence, the model is fully specified by the block assignments of nodes $\{g_i\}$ and the mixing matrix $\Omega=\{\omega_{rs}\}$ governing the probabilities of observing one edge between each pairs of nodes from blocks $r$ and $s$. If the diagonal elements $\omega_{rr}$ in the mixing matrix are larger than the off-diagonal elements, then the networks generated by the stochastic block model have the traditional assortative communities. When the off-diagonal elements $\omega_{rs}$ for $r \neq s$ are larger than the diagonal elements, the generated network contains disassortative mixing patterns, such as the structures observed in the core-periphery graphs~\cite{borgatti2000models} and the bi-partite graphs. In general, the inference of the stochastic block model aims at discovering the mixing matrix and nodes block assignment which maximize the likelihood of generating the observed network. It does not impose any constraint on which type of structure is discovered. Thus, the inference of the stochastic block model is more general than the traditional modularity maximization because it may recover a set of assortative or disassortative structures from the observed network.

Given the nodes' block assignments of the standard stochastic block model, the nodes in the same block are by definition statistically indistinguishable, and they all have the same expected node degrees. Therefore, the most likely block assignments often cluster the nodes of similar degrees into a block, resulting in clusters with homogeneous node degrees, rather than heterogeneous node degrees in each block, typical for the traditional community structures. The degree-corrected stochastic block model~\cite{karrer2011stochastic} defines the expected number of edges between nodes $i$ and $j$ as $\lambda_{ij} = \omega_{g_i,g_j} \beta_i \beta_j$ where for each node $l$, $\beta_l$ is a model parameter associated with it. Note that $\lambda_{ij}$ is defined as the expected number of edges and not as customary the probability, assuming multi-edges (multiple edges between a pair of nodes) are allowed in this definition~\cite{karrer2011stochastic}. The expected node degrees in the degree-corrected stochastic block model, given the maximum likelihood estimates of these model parameters, are equal to the degrees observed in the input network. Hence, the degree-corrected stochastic block model allows a community to have a wide range of node degree distribution. 

The simple yet effective degree-correction modification improves the performance of statistical inference of community structure in complex networks. When the nodes in an assortative community have a wide range of degree distribution, the statistical inference of the standard stochastic block model is likely to detect the sub-optimal disassortative structures, while the statistical inference of the degree-corrected model is more likely to discover the sought-after assortative structures. Yet, the degree-corrected stochastic block model does not impose any constraints on the mixing pattern of the resulting block assignments, thus the return of the assortative structures is not guaranteed.

As suggested in~\cite{peel2017ground}, both the traditional assortative community structures and the other disassortative structures are potentially good fits to the degree-corrected stochastic block models. Hence, depending on the starting point, the inference algorithm like Markov chain Monte Carlo tends to get trapped in the local optimum of the log-likelihood. Therefore, in this paper, we propose a regularized stochastic block model which provides an extra regularization term to guide the discovery of assortative or disassortative structure by the statistical inference of the stochastic block model. Unlike the modularity maximization algorithm which always attempts to find traditional assortative communities, the inference of this regularized stochastic block model controls the mixing patterns discovered in the given network.

\section*{Results}
\subsection*{Degree-corrected stochastic block model}
The degree-corrected stochastic block model~\cite{karrer2011stochastic} is a generative model of the graph in which the edges are randomly placed between nodes. The probability of observing an edge between two nodes depends exclusively on the block assignments of its two endpoints. In the degree-corrected stochastic block model, the nodes in the same block are statistically indistinguishable from each other. The degree-corrected stochastic block model assumes the number of edges between nodes $i$ and $j$ is independently distributed Poisson random variable with the mean $\lambda_{g_i,g_j}$. 

Let $\bf A$ be the adjacency matrix whose component $A_{ij}$ denotes the number of edges between nodes $i$ and $j$ in an unweighted undirected multigraph. Multi-edges (multiple edges between a par of nodes) and self-loop edges (the edges connecting a node to itself) are practical in certain networks such as the web network where a web page may contain multiple hyperlinks pointing to other pages and to itself. Such edges are less common in social networks. However, most social networks are very sparse, so the impact of multi-edges and self-loop edges is negligible in such networks. Let a node $i$ with $k$ self-loop edges be represented by the diagonal adjacency matrix element $A_{ii} = 2k$ and the number of edges between two different nodes $i$ and $j$ follows the Poisson distribution with mean $\lambda_{ij}$ while the number of self-loop edges at node $l$ follows the Poisson distribution with mean $\frac{1}{2}\lambda_{ll}$. 

Given the parameters $\{\lambda_{ij}\}$, the likelihood of generating $\bf A$ is
\begin{equation} \label{eq:ll0}
    \log P({\bf A}|\{\lambda_{ij}\}) = \prod_{i} \frac{(\frac{1}{2}\lambda_{ii})^{A_{ii}/2} }{(\frac{1}{2}A_{ii})!} e^{-\lambda_{ii}/2} \prod_{i<j} \frac{\lambda_{ij}^{A_{ij}} }{A_{ij}!} e^{- \lambda_{ij}}.
\end{equation}
Note that $\lambda_{ij}$ is defined as the number of edges here and not as customary the probability, because multi-edges are allowed in this definition. In an unweighted undirect network, after ignoring all terms independent of the $\lambda_{ij}$s, the log-likelihood above simplifies to
\begin{equation} \label{eq:ll}
    \log P({\bf A}|\{\lambda_{ij}\}) = \frac{1}{2} \sum_{ij} \Big(A_{ij} \log \lambda_{ij} - \lambda_{ij} \Big).
\end{equation}
In the degree-corrected stochastic block model, the model parameter $\lambda_{ij}$ is defined as $\lambda_{ij} = \omega_{g_i,g_j} \beta_i \beta_j$ where for each node $l$, $\beta_l$ is a parameter associated with this node. Given a partition of the network, i.e., its block assignments $\{g_i\}$, reference~\cite{karrer2011stochastic} obtains the maximum likelihood estimates of the model parameters as $\hat{\beta}_i = \frac{k_i}{\kappa_{g_i}}$ and $\hat{\omega}_{rs} = m_{rs}$, where $k_i$ is the degree of node $i$, $\kappa_{r}$ is the sum of the degrees of all nodes in a block $r$, and $m_{rs}$ is the total number of edges between blocks $r$ and $s$, or, if $r=s$, twice the number of edges in $r$. Note that each parameter $\omega_{rs}$ in the degree-corrected stochastic block model is not the probability of observing an edge between endpoints in blocks $r$ and $s$. Instead, it is a model parameters that depends exclusively on the block assignments $r$ and $s$.

As Eq.~\ref{eq:ll} indicates, the degree-corrected stochastic block model does not specify the mixing pattern of the resulting block assignments. It is possible that the Poisson means $\lambda_{ij}$ for the pairs of nodes in the same block are much smaller than the corresponding value between different blocks, which corresponds to the disassortative mixing pattern. Hence, the disassortative mixing pattern can also be a good fit to the degree-corrected stochastic block model. This observation inspires us to propose a regularization term for the degree-corrected stochastic block model, so it controls the mixing patterns discovered in the given network.

\subsection*{Assortative and disassortative structures}

As defined in previous literatures~\cite{newman2004finding,newman2003structure,fortunato2010community}, the assortative structures correspond to the traditional community structures, nodes are more frequently connected with each other inside communities than with the rest of the network. The disasortative structures do not satisfy this condition. For example, the core-periphery structures~\cite{borgatti2000models} divides the networks nodes into the core part, where nodes are often the hubs of the networks, and the periphery part, where nodes of low-degree connect to the core nodes. This is an advantage of the degree-corrected stochastic block model because it allows the discovery of the structures with different mixing patterns. However, when searching for the weak assortative community structures, the inference algorithm may still return the disassortative structure\cite{decelle2011asymptotic} instead.

We generate synthetic networks using the degree-corrected stochastic block model with a block assignment $\{g_i\}$ and the parameters $\omega_{rs}$ chosen for the assortative communities as follows
\begin{equation} \label{eq:synthetic}
\omega_{rs}=
    \begin{cases}
    &\gamma \omega_{0} \quad \quad  \text{ if  } r=s,\\
    &\omega_{0} \quad \quad \quad  \text{if  } r\neq s.
    \end{cases}
\end{equation}
where a large value of $\gamma$ results in strong community structure while $\omega_{0}$ controls the sparsity of the network. The degree sequence is drawn from a power-law distribution with exponent $2.5$, and the block assignments are randomly assigned, but each block is of the same size.

\begin{figure*}[!htb]
    \centering
    \subfloat[Local optima of log-likelihood]{\includegraphics[width=9cm]{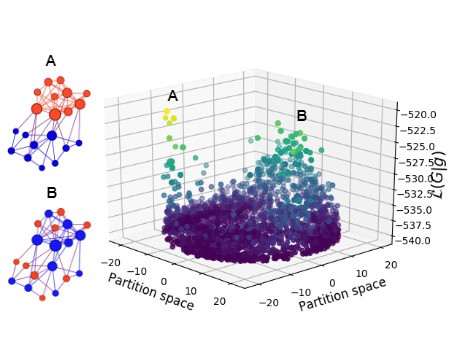}}
    \subfloat[Convergence]{\includegraphics[width=7cm]{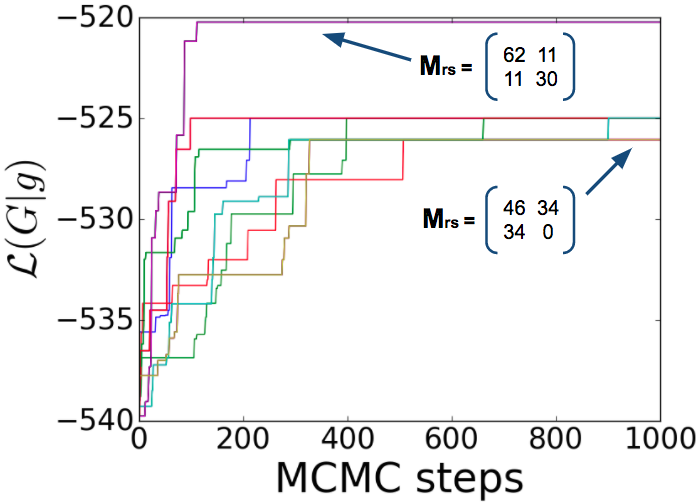}}
    \caption{The convergence of 20 Markov Chain Monte Carlo (MCMC) trials for the degree-corrected stochastic block model. The local optimum $A$ found by MCMC represents the assortative communities, whereas there are other local optima representing disassortative structures such as point B. \textbf{(a)} Multiple locally optimal partitions discovered by the MCMC inference for degree-corrected stochastic block model; \textbf{(b)} 2 out of the 20 MCMC trials find the most likely and sought-after assortative partition, while the other trials get trapped at the local optima. The matrix $M_{rs}$ indicates the number of edges between every pair of blocks.}
    \label{fig:dcsbm_3dscatter}
\end{figure*}

Given the synthetic networks produced by the degree-corrected stochastic block model, we infer the block assignment $\{g_i\}$, recovering the parameters used for its generation. Specifically, the number of communities is set as two for both the generation and recovery. To generate samples, the degree-corrected stochastic block model uses $\omega_{0}=0.01$ and $\gamma=10$. Each of the two blocks contains $10$ nodes. We scatter the sampled partitions on the x-y plane in Figure~\ref{fig:dcsbm_3dscatter} and the z-axis indicates the log-likelihood of the corresponding partitions. We adapt the Markov chain Monte Carlo (MCMC) algorithm~\cite{nasrabadi2007pattern,peixoto2014efficient} to infer the block assignment of the stochastic block model and its extensions. The MCMC algorithm~\cite{peixoto2014efficient} samples the distribution of all possible block assignments in the network partitions space. For each trail, the algorithm starts from a random partition and iteratively, with certain probability, traverse to an adjacent partition. An entire MCMC sweep of all nodes in the network requires $O(E)$ operations where $E$ is the number of edges in the networks~\cite{peixoto2014efficient}. The mixing time of the MCMC algorithm depends on the specific input network and the starting point. The details of the MCMC algorithm are discussed in the Supplementary Information.

Figure~\ref{fig:dcsbm_3dscatter} shows the existence of multiple local optima in the log-likelihood of degree-corrected stochastic block model. The inference process finds three local optima here: (i) partition A which corresponds to the assortative structure matching the ground truth block assignment used for its generation; (ii) partition B which corresponds to a disassortative structure; and (iii) another disassortative partition which is not explicitly marked in the Figure~\ref{fig:dcsbm_3dscatter}(a). Under the degree-corrected stochastic block model, the MCMC inference starting from random initial partition finds the most suitable partition A in only 2 out of 20 trials, whereas the other attempts are trapped at the local optima as shown in Figure~\ref{fig:dcsbm_3dscatter}(b).

Since there are multiple local optima in the log-likelihood of the degree-corrected stochastic block model, the inference algorithm may converge to any of them nondeterministically. Specifically, the type of the discovered structure depends on the trial starting point and inference algorithm parameters. To avoid such nondeterministic outcomes, we introduce a novel approach called Regularized Stochastic Block Model (RSBM) applicable to any inference algorithm. RSBM constrains nodes' internal degree ratios, each of which is defined in the objective function as the fraction of a node's neighbors inside the same community. The resulting algorithm reliably finds assortative or disassortive structures as directed by the value of a single parameter. 

\subsection*{Regularized stochastic block model}

We extend the formulation of the expected number of edges between nodes $i$ and $j$, determined by the \textit{Poisson} rate $\lambda_{ij}$, in the degree-corrected stochastic block model by defining it as
\begin{equation} \label{eq:theta}
    \lambda_{ij} = 
    \begin{cases}  \omega_{g_i,g_j} I_i I_j    & \mbox{if} \quad g_i=g_j\\
                                 \omega_{g_i,g_j} O_i O_j  & \mbox{otherwise}
    \end{cases}
\end{equation}
where any node $l$ has two associated parameters $I_l$ and $O_l$. Given Eq.~\ref{eq:ll}, the log-likelihood of generating graph $G$ by this regularized stochastic block model can be written as
\begin{align} \label{eq:logl_IO}
    \mathcal{L}(G|{\bf g,\omega, I, O})  = 2\sum_i \Big( k_i^+ \log I_i + k_i^- \log O_i \Big) + \sum_{rs} m_{rs} \log \omega_{rs} - \omega_{rs} \Lambda_{rs} 
\end{align}
where $k_i^+$ is the number of neighbors of node $i$ which are inside the same block given the block assignment $\bf g$ and $k_i^- = k_i - k_i^+$. Thus, we get  
\begin{align}
   \Lambda_{rs} = \begin{cases}  (\sum_{i \in r} I_i)^2   & \mbox{if } r=s \\ 
 \sum_{i \in r} O_{i}  \sum_{i \in s} O_{i} & \mbox{if } r\neq s \end{cases} 
\end{align}
To simplify, we write $i\in r$ if $g_i = r$. For block assignment $\bf g$, the maximum-likelihood values of $\omega_{rs}$ are
\begin{equation} \label{eq:mle_omega}
    \hat{\omega}_{rs} = \frac{m_{rs}}{\Lambda_{rs}}
\end{equation}
Dropping the constants and substituting using Eq.~\ref{eq:logl_IO}, we obtain
\begin{align} \label{eq:important}
    \mathcal{L}(A|{\bf g,I,O})
    &=  \sum_{rs} m_{rs} \log \frac{m_{rs}}{\Lambda_{rs}} + 2\sum_i \Big( k_i^+ \log I_i + k_i^- \log O_i \Big) 
\end{align}

Note that if we set $I_i = O_i = 1$ here, the log-likelihood above reduces to the definition of standard stochastic block model with $\Lambda_{rs} = n_r n_s$ which is exactly the product of the sizes of two blocks $r$ and $s$. When $I_i = O_i = k_i$, the second sum on the right hand side (RHS) becomes irrelevant to the maximum likelihood estimation (MLE) result. Hence, the log-likelihood reduces to the definition of degree-corrected stochastic block model in Eq.~\ref{eq:important} with $\Lambda_{rs} = \kappa_r \kappa_s$, i.e., the product of the sums of degrees of nodes in two blocks $r$ and $s$. Hence, by introducing two sets of parameters ${\bf I} = \{I_i\}$ and ${\bf O} = \{O_i\}$ in the edge probability, we obtain a more generalized definition of stochastic block model here.

\subsection*{Regularization by prior in-degree ratios }
For alternative formulation of our model, we define the prior in-degree ratio $f_i = I_i / (I_i + O_i)$ and $\theta_i = I_i + O_i$ for each node $i$. By rewriting the second summation on the RHS of Eq.~\ref{eq:important}, we get 
\begin{align} \label{eq:important2}
    \mathcal{L}(G|{\bf g,I,O})
    &=  \sum_{rs} m_{rs} \log \frac{m_{rs}}{\Lambda_{rs}} - 2\sum_i k_i H(\frac{k_i^+}{k_i},f_i) + 2 \sum_i k_i \log \theta_i
\end{align}
where $H(\frac{k_i^+}{k_i},f_i) = -\frac{k_i^+}{k_i} log f_i - \frac{k_i^-}{k_i} log (1 - f_i)$ represents the cross entropy between the observed and prior in-degree ratio. Therefore, the prior in-degree ratios $\{f_i\}$ regularizes the in-degree ratios $\{\frac{k_i^+}{k_i}\}$ in the resulting partition by maximizing Eq.~\ref{eq:important2}.

In real networks, the low degree nodes are more likely to have neighbors inside a block than the high degree nodes are. Suppose $f_i$ depends only on the degree of node $i$, i.e. $f_i = f(k_i)$. Then, the function $f(k): \mathbb Z_+ \to [0, 1] $ should be strictly decreasing. In an assortative partition of the network, we have
\begin{itemize}
    \item $f(1) = 1$ because a node with degree one must connect to the community it belongs to;
    \item for $k\approx |V|$, $f(k) \ll 1$ because a super-hub eventually does not belong to any community as its degree is of the order of the number of nodes in the entire network.
\end{itemize} 
A simple function $\{f(k)\}$ satisfying this requirement is of the form
\begin{equation}
  f(k) = \alpha + \frac{(1 - \alpha)}{k}
\end{equation}
where $\alpha$ is the only extra parameter we introduce to the regularized stochastic block model (RSBM). Alternatively, we can select a constant $f\in(0,1)$  such that  
\begin{equation}
    f(k)= max(f, \frac{1}{k})
\end{equation}
The impact of different choices of $f_i$ on the discovered block assignment is discussed in the following two subsections presenting experimental results. It is worth noting that, for either choice of the prior in-degree ratio $\{f_i\}$, there is only one extra parameter, i.e. $\alpha$ or $f$, introduced here in addition to the original parameters of the degree-corrected stochastic block model.

\subsection*{Experimental Results}
We generate synthetic networks with the assortative communities using the degree-corrected stochastic block model. The parameters $\omega_{rs}$ in the network generation process are specified by Eq.~\ref{eq:synthetic}. By selecting a small value of $\gamma > 1$ in Eq.~\ref{eq:synthetic}, the generated community structures are relatively weak, thus, it is more difficult to detect them using the statistical inference of the degree-corrected stochastic block model.

\begin{figure*}[!htb]
    \centering
    \subfloat[Local optimum of log-likelihood]{\includegraphics[width=9cm]{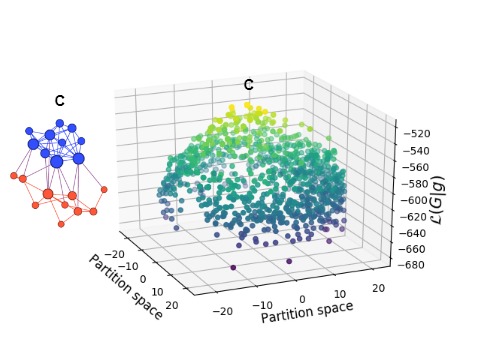}}
    \subfloat[Convergence]{\includegraphics[width=7cm]{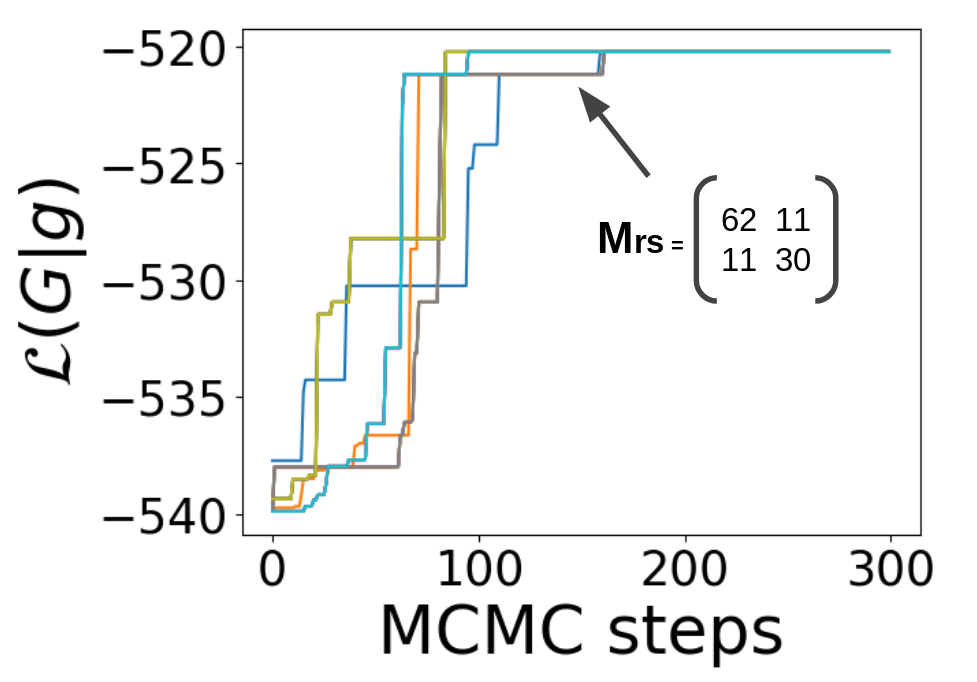}}
    \caption{The convergence of 20 Markov Chain Monte Carlo (MCMC) trials for the regularized stochastic block model (RSBM) introduced here. All trials converge to the local optimum C which represents an assortative structure. \textbf{(a)} One local optimal partition observed during the MCMC inference for our model. \textbf{(b)} All twenty MCMC trials find the sought-after assortative structure.}
    \label{fig:gsbm_3dscatter}
\end{figure*}

We adapt the Markov chain Monte Carlo (MCMC) algorithm~\cite{nasrabadi2007pattern} to infer the block assignment of the stochastic block model and its extensions. Figure~\ref{fig:gsbm_3dscatter}(a) shows there is only one unique local optimum, the partition C, found by 20 MCMC trials under the regularized stochastic block model. Therefore, all 20 MCMC trials converge to this unique local optimum. As shown in Figure~\ref{fig:gsbm_3dscatter}(b), the Markov Chain Monte Carlo inference finds the correct block assignment within only 150 steps. The regularization terms made it possible to avoid the unsuitable local optima during the inference.

\begin{figure*}[!htb]
    \centering
    \subfloat[$f=0.14$]{\includegraphics[width=5cm]{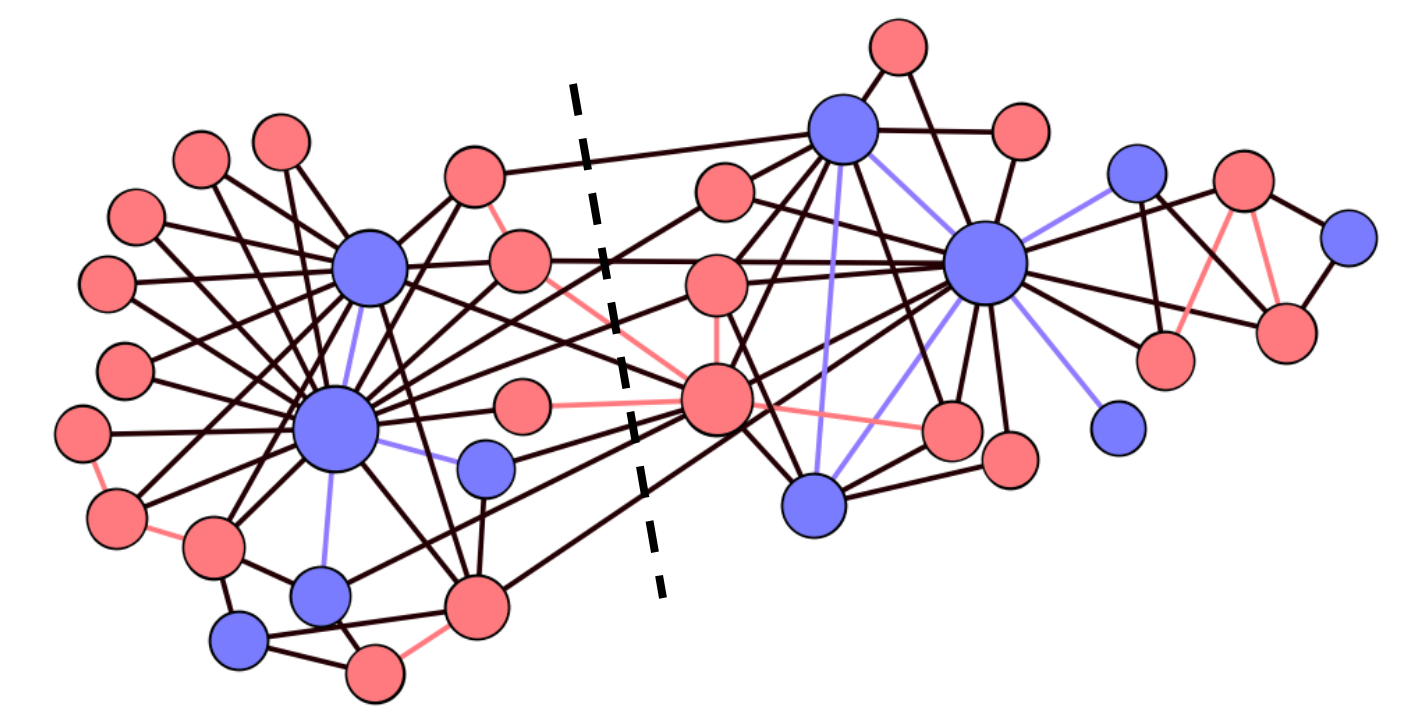}}
    \subfloat[$f=0.35$]{\includegraphics[width=5cm]{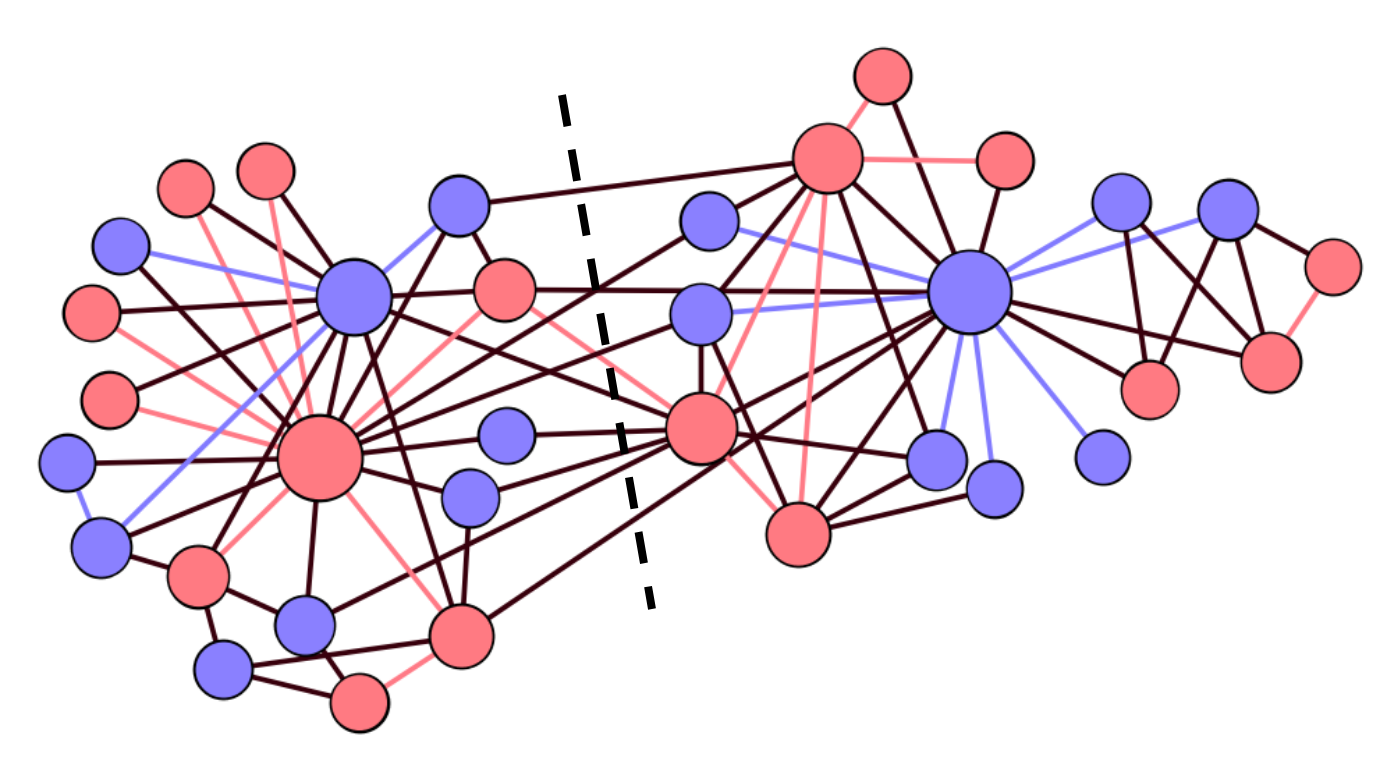}}
    \subfloat[$f=0.56$]{\includegraphics[width=5cm]{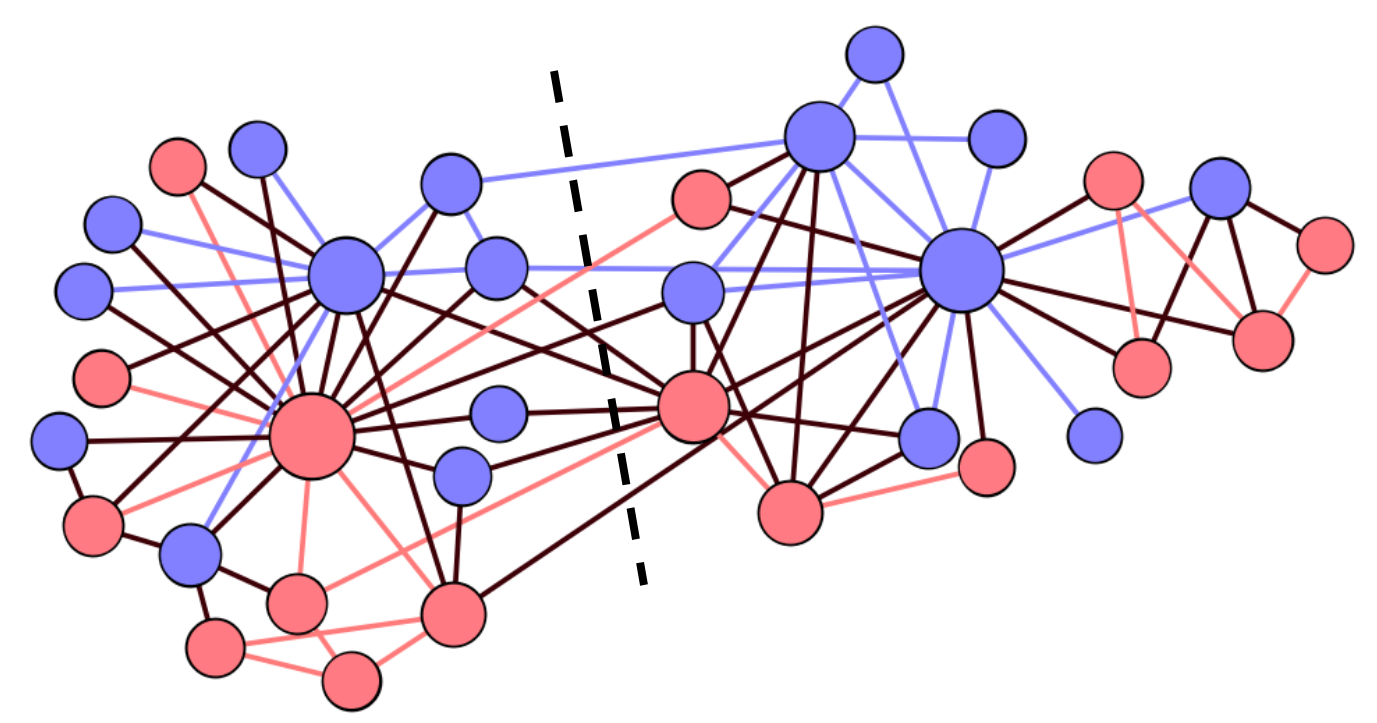}} \\
    \subfloat[$f=0.68$]{\includegraphics[width=5cm]{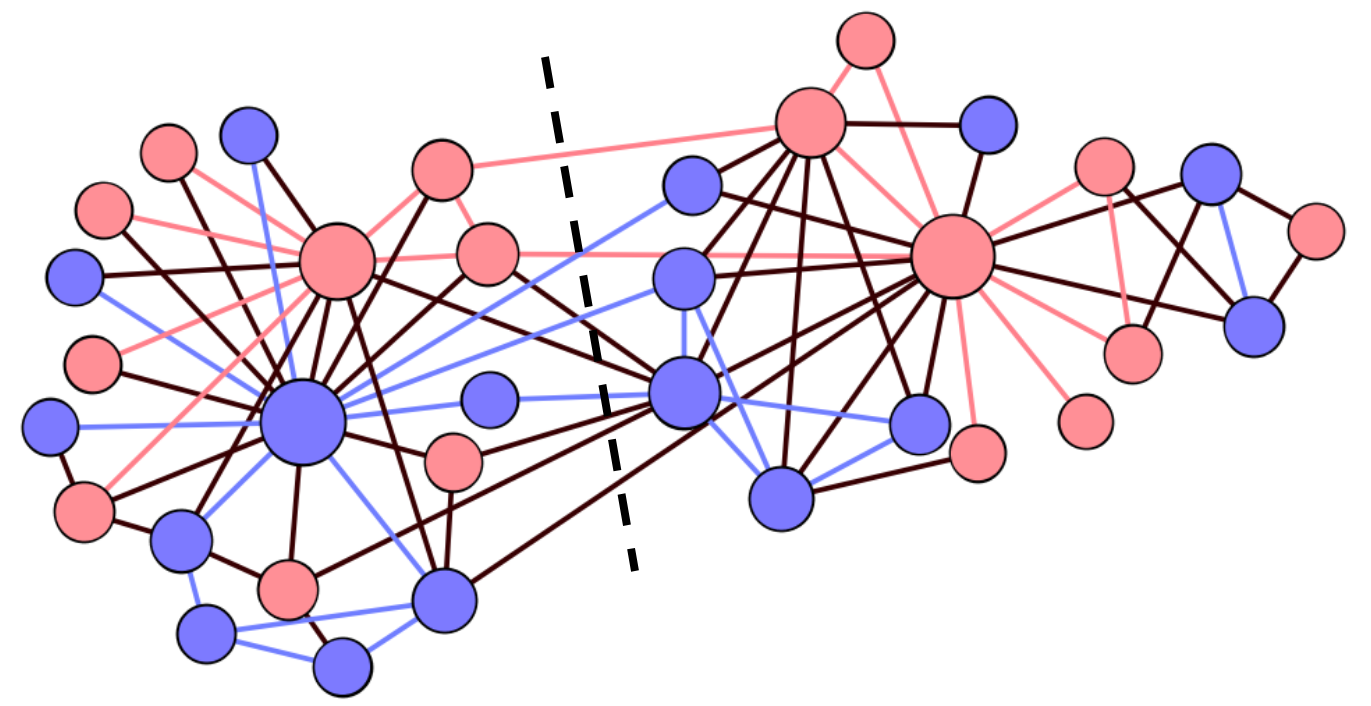}}
    \subfloat[$f=0.81$]{\includegraphics[width=5cm]{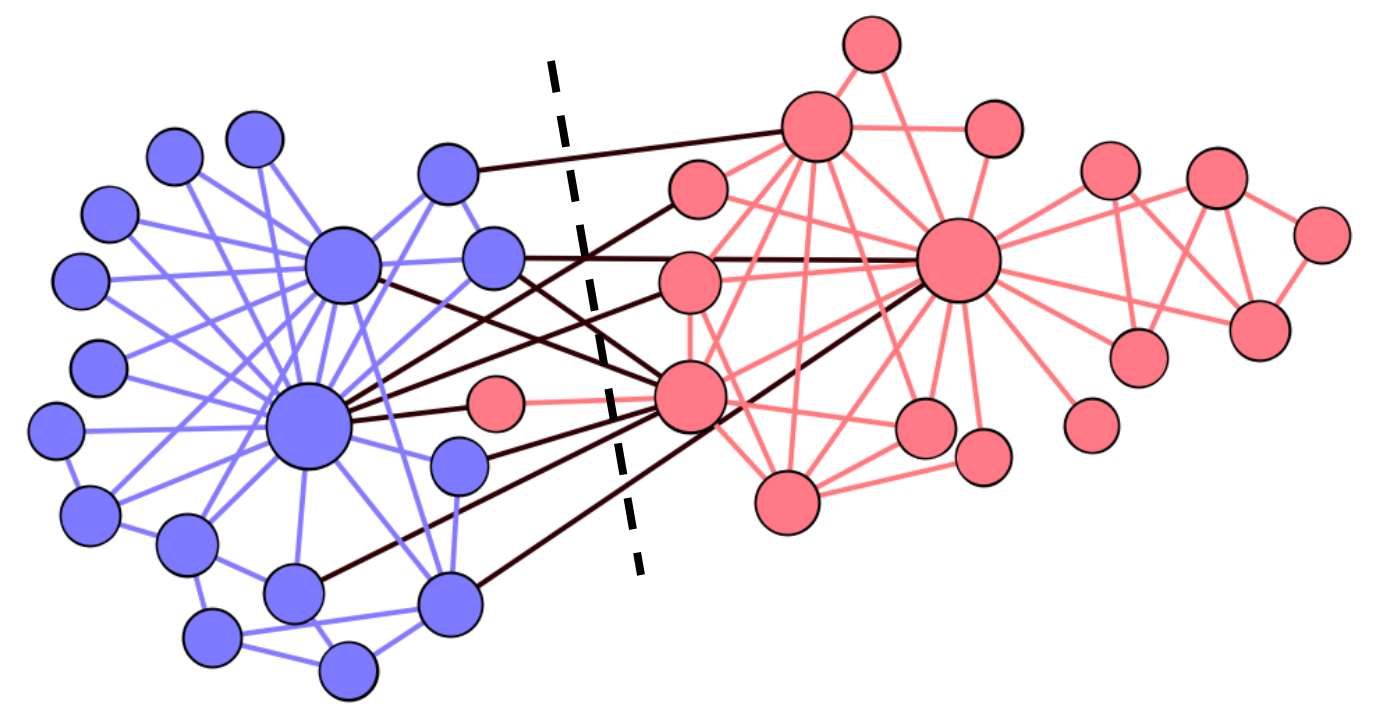}}
    \caption{The partition of Karate network inferred by Markov Chain Monte Carlo under different parameter settings where nodes of the same color belong to the same partition. The black dotted line represents the ideal partition in the ground truth. A small $f$ results in the core-periphery partition of the network while a large $f$ leads to assortative partitions. The values of the $f$ parameter are shown in the corresponding sub-figures captions.}
    \label{fig:karate_figure}
\end{figure*}

We use the networks including Karate club network of Zachary~\cite{zachary1977information}, the Dolphin social network of Lusseau et al.~\cite{lusseau2003bottlenose} and the network of fictional characters' interactions in the novel Les Miserables by Victor Hugo~\cite{newman2004finding} to demonstrate the performance of the regularized model introduced here. The details of each network are presented in the Supplementary Materials. For Karate club network, we evaluate the effect of the regularization terms on the resulting partitions using Markov chain Monte Carlo (MCMC) as the inference algorithm. For every node $i$ in the network, we set the parameter $\theta_i = k_i$ and $f_i = max(f, 1/k_i)$ for our regularized stochastic block model defined by Eq.~\ref{eq:important2} where $k_i$ is the degree of node $i$ and $f_i$ represents the prior in-degree ratio for regularization. Figure~\ref{fig:karate_figure} shows the most likely partition of the Karate club network found by MCMC using different $f$ values. The color represents the block assignment, and the black dashed line divides the network into two parts in the ground truth partition. As shown in Figure~\ref{fig:karate_figure}, when $f=0.14$, the inference algorithm outputs a core-periphery structure which clusters high-degree nodes into the blue block and the remaining low-degree nodes into the red block. This is because the sum of cross entropy terms serves as a regularization term which penalizes those partitions that assign adjacent nodes into the same block. As the value of $f$ grows, the inference algorithm is more likely to detect assortative structure. When $f=0.85$, the inferred block assignment matches the ground truth partition of the Karate club network with the exception of one single red node. However, this node has only one connection to each block; thus, it is quite arguable to which block this node should belong.

\begin{figure*}[!htb]
    \centering
    \subfloat[Karate club]{\includegraphics[width=6cm]{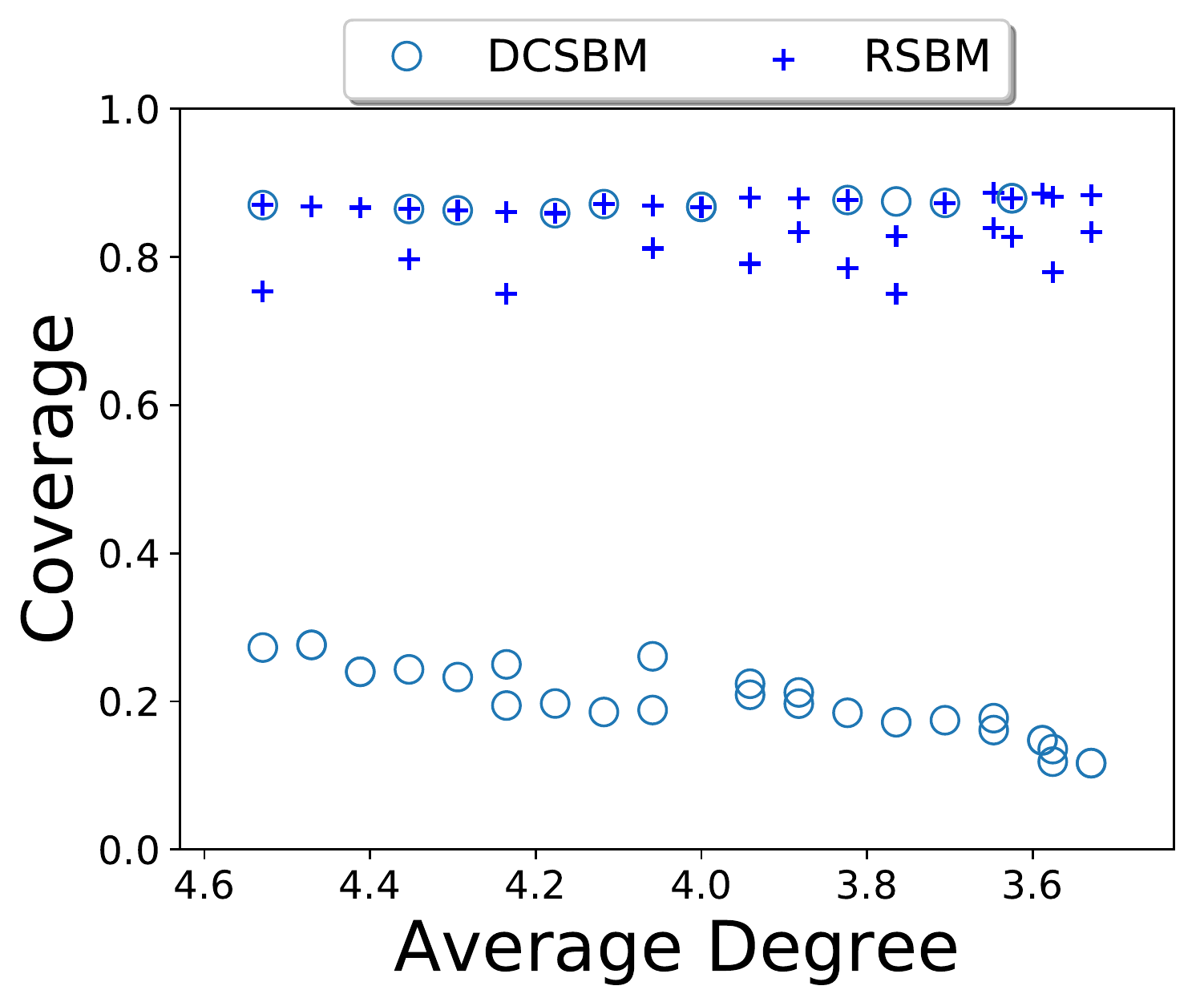}}
    \subfloat[Dolphin social]{\includegraphics[width=6cm]{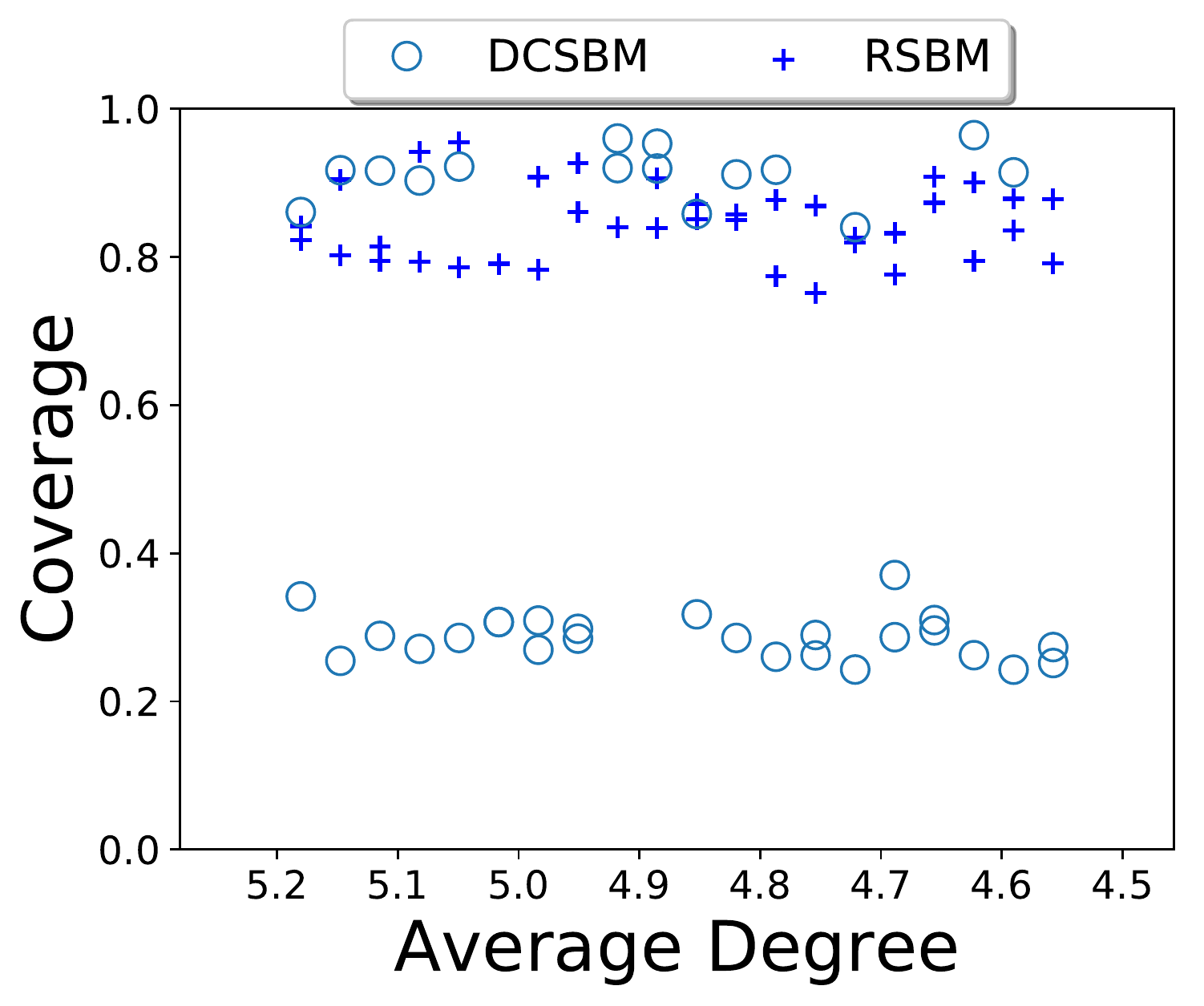}} 
    \subfloat[Les Miserables]{\includegraphics[width=6cm]{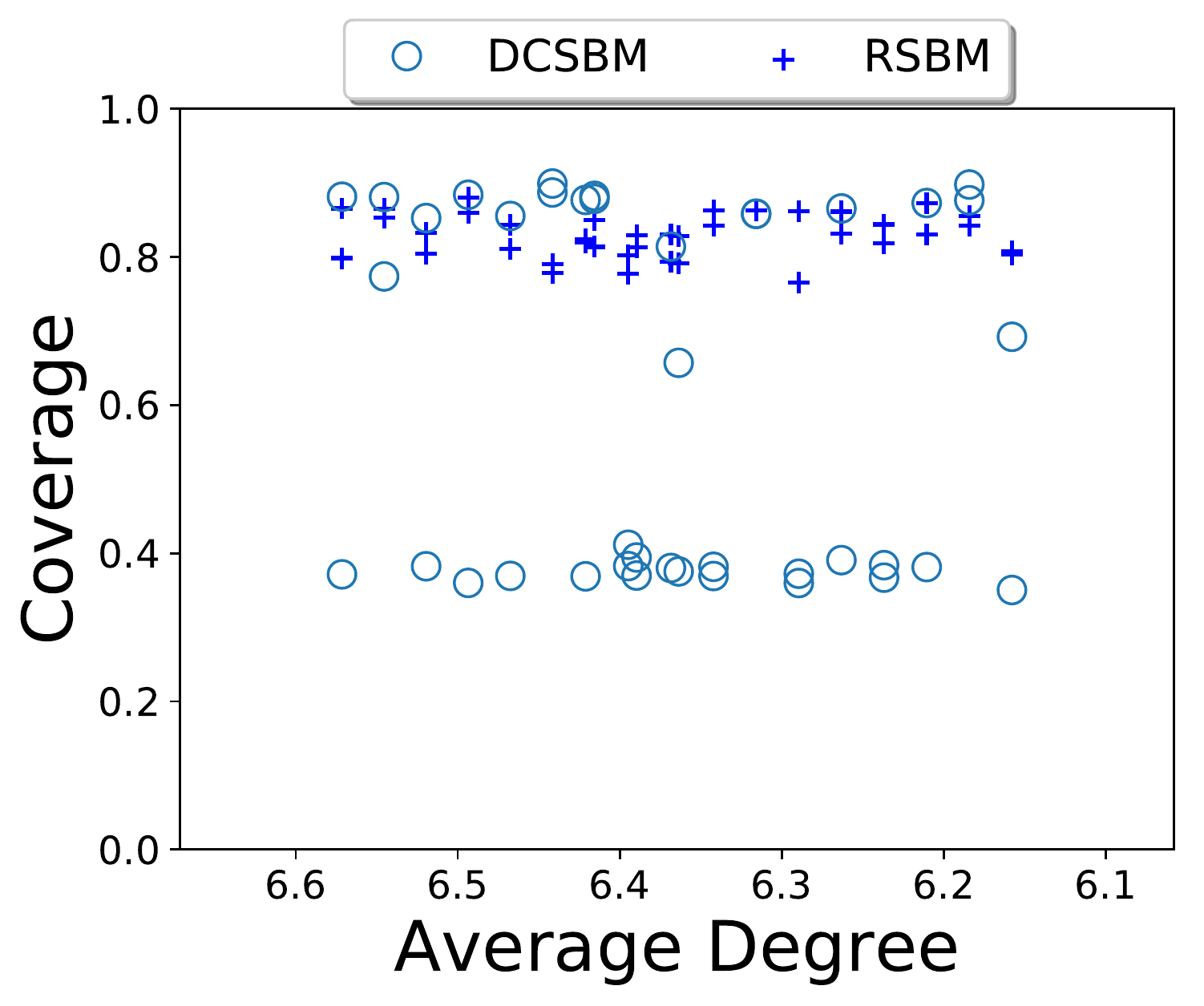}}
    \caption{The \textit{coverage} of partitions in real networks as a function of average node degree. The optimal partitions are inferred by the Markov Chain Monte Carlo algorithm under degree-corrected stochastic block model (DCSBM), marked by circles, and its regularized extension (RSBM) introduce here, marked by crosses. \textbf{(a)} Karate club network~\cite{zachary1977information}; \textbf{(b)} Dolphin social network~\cite{lusseau2003bottlenose}; \textbf{(c)} Characters interaction network of Les Miserables~\cite{newman2004finding}}
    \label{fig:gsbm_real}
\end{figure*}

The results in Figure~\ref{fig:gsbm_real} show that the MCMC inference of our RSBM model on these networks finds partitions with higher \textit{coverage} than the ones inferred by the degree-corrected stochastic block model. The \textit{coverage} of a partition~\cite{fortunato2010community} is defined as the ratio of the number of edges with both endpoints in the same block to the total number of edges in the entire network
\begin{equation}
    \text{coverage({\bf g})} = \frac{|\{(i,j)\in E| g_i = g_j\}|}{|E|}
\end{equation}
A low \textit{coverage} indicates that the resulting partition is disassortative. An ideal assortative partition of the network, where all clusters are disconnected, yields a \textit{coverage} of 1.

Figure~\ref{fig:gsbm_real} shows the \textit{coverages} of the partitions found in the three real networks mentioned above. We randomly remove edges in these networks to further increase the sparsity of these networks. And the numbers of blocks used in trials are set to their values for these real networks well-accepted in the literature. The results in Figure~\ref{fig:gsbm_real} indicate that, under degree-corrected stochastic block model (DCSBM), the inference algorithm is likely to miss the assortative structures and returns instead the disassortative partitions of the network, which also fit the model in such cases. In contrast, the MCMC inference of our regularized stochastic block model (RSBM) almost deterministically produces the assortative structures. Interestingly, the inference of degree-corrected stochastic block model produces the partitions with the \textit{coverage} values distributed at two levels. There is no partition with a \textit{coverage} between these two levels found by the inference algorithm. This is similar to the case of the synthetic network in Figure~\ref{fig:dcsbm_3dscatter} where both the assortative communities and disassortative structures fit the degree-corrected stochastic block model. Which structure is found is determined by random sampling of the potential partitions. In other words, there is no way to guide whether the disassortative structures or the assortative communities are preferred, so, two consecutive MCMC trials may return completely different structures.

In contrast, the inference of the introduced here regularized stochastic block model using the prior in-degree ratio $f_i = 0.8 + 0.2 / k_i$ is very robust. It only produces partitions with a high \textit{coverage}, which are generally distributed at the same level as the assortative partitions under degree-corrected stochastic block model. Moreover, the sparsity of the networks does not have an obvious impact on the resulting partitions under our regularized model.

\begin{figure*}[!htb]
    \centering
    \subfloat[Karate club]{\includegraphics[width=5cm]{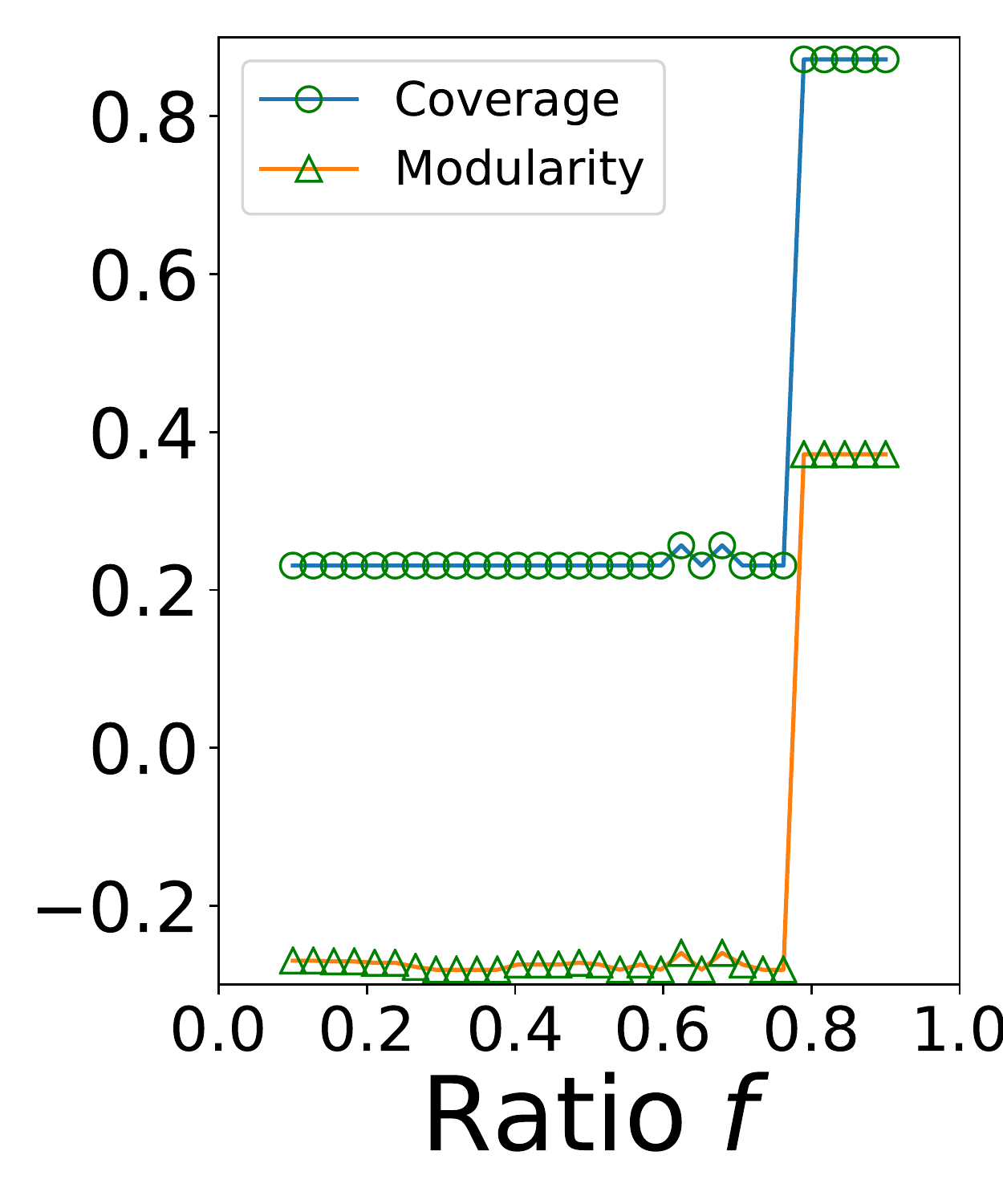}} 
    \subfloat[Dolphin social]{\includegraphics[width=5cm]{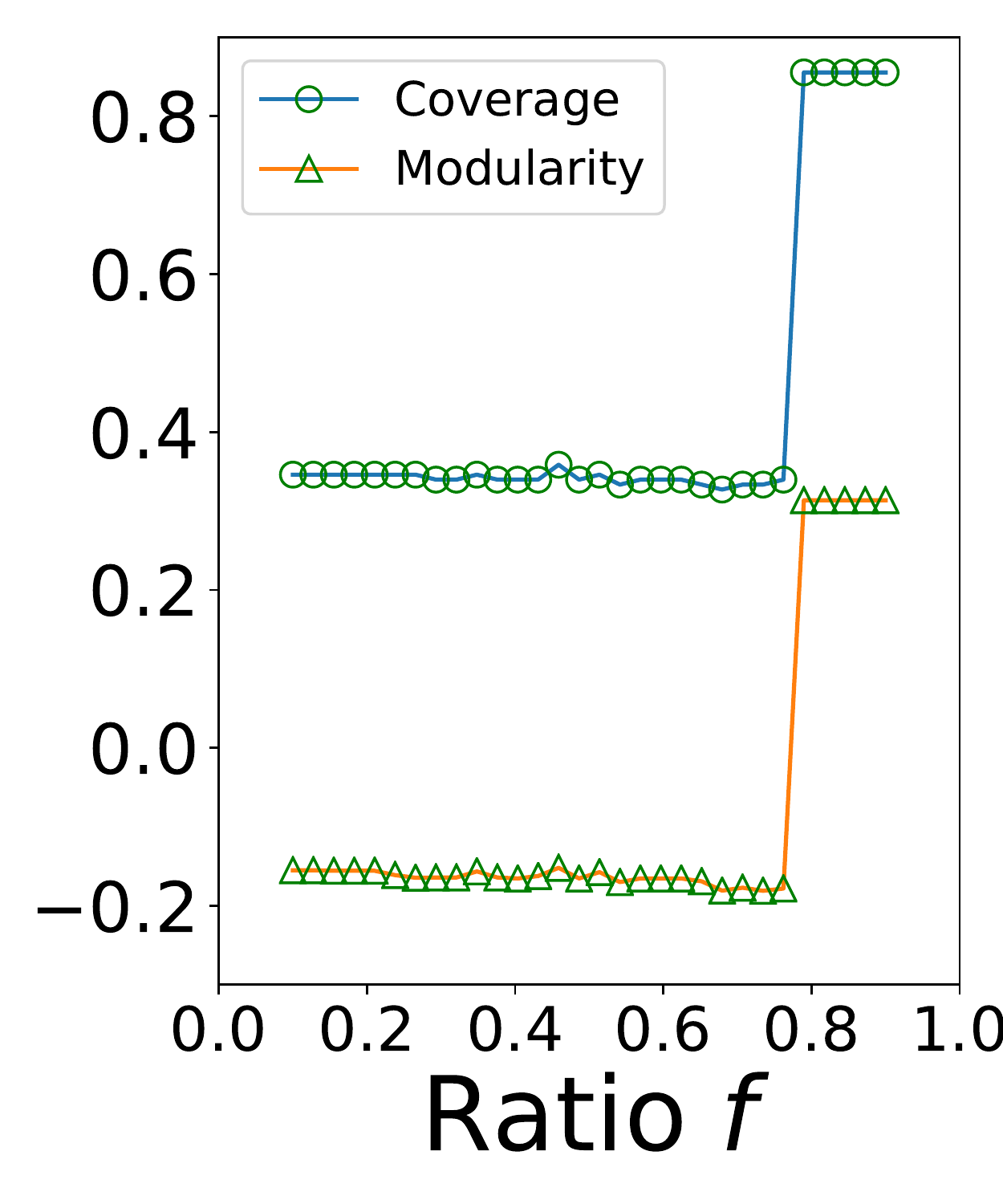}} 
    \subfloat[Les Miserables]{\includegraphics[width=5cm]{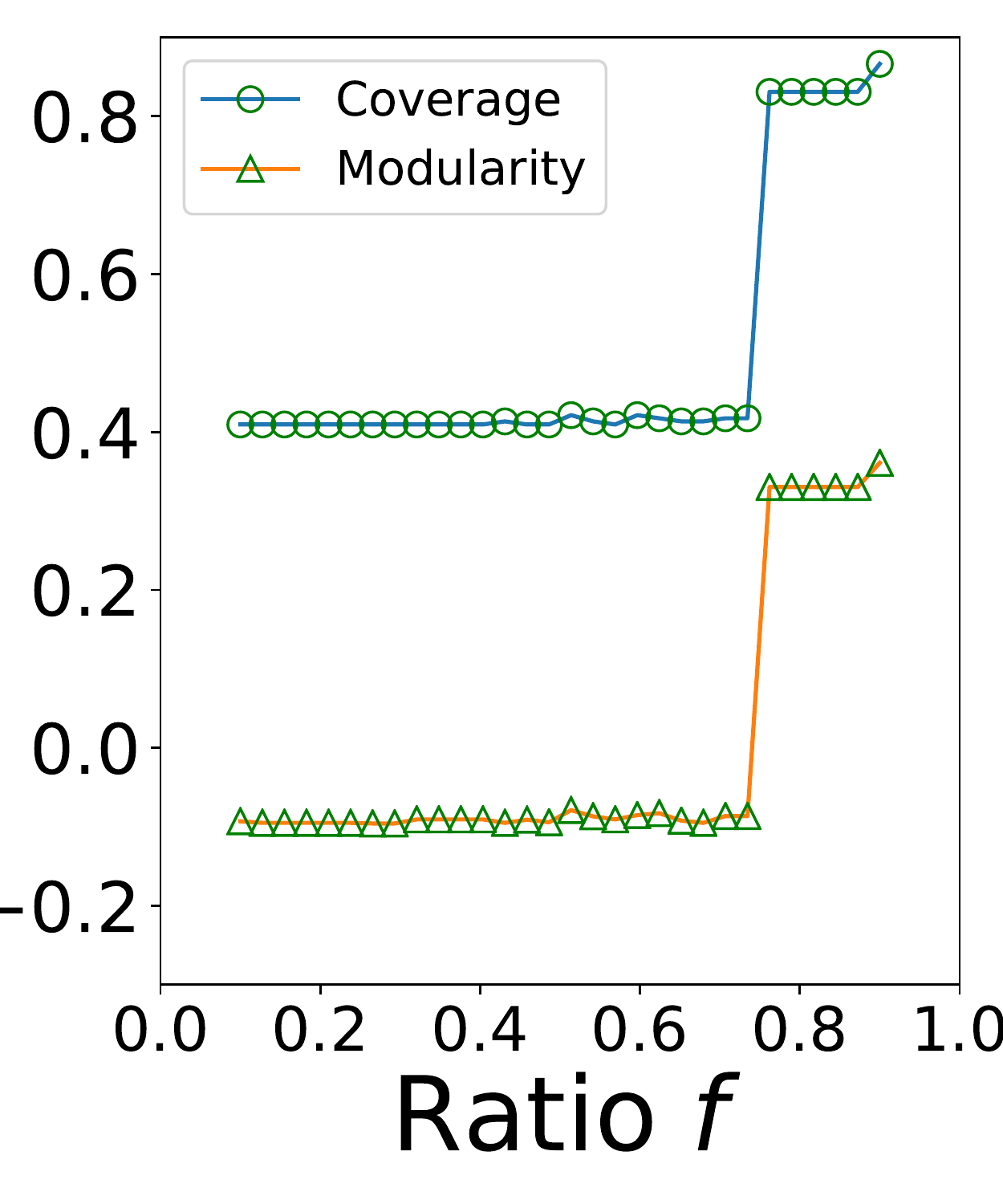}}
    \caption{The \textit{coverage} and modularity  of partitions in real networks as a function of parameter $f$. The optimal partitions are inferred by Markov Chain Monte Carlo algorithm under degree-corrected stochastic block model (DCSBM) and its regularized extension (RSBM) introduced here. The networks used for evaluation include \textbf{(a)} Karate club network~\cite{zachary1977information}; \textbf{(b)} Dolphin social network~\cite{lusseau2003bottlenose}; \textbf{(c)} Characters interaction network of Les Miserables~\cite{newman2004finding}}.
    \label{fig:gsbm_real2}
\end{figure*}

We evaluate the modularity~\cite{newman2006modularity} of resulting partitions of the real networks as a function of parameter $f$. 
A high modularity indicates the strong community structure. In our experiments, we use a constant $f$ such that $f_i=f$ for each
node $i$ in the network with degree $k_i>1/f$, otherwise $f_i=1/k_i$. We start with a small $f$ value and increase it in each iteration.
The MCMC inference uses as a starting point the original network initially, and the network partition found in the previous iteration subsequently. Figure~\ref{fig:gsbm_real2} shows that, as the value of $f$ increases, in general both the modularity and the \textit{coverage} grow. When $f$ is close to 0, the value of $f$ does not have any effect on the network partition. However, as $f$ becomes larger, then at a certain threshold of about 0.75, the increase of $f$ leads to a critical transition from disassortative partitions to assortative communities. Then, a larger $f$ value does not further increase the modularity  of the partition. These results indicate that, with one single parameter $f$, the Markov Chain Monte Carlo inference gains the flexibility to choose between assortative communities resulting from clustering dense modules of nodes together or disassortative structure, such as the core-periphery structure from clustering high-degree and low-degree nodes into separate blocks. With a choice of high $f$, the inference algorithm is likely to detect assortative communities.

\section*{Discussion}
The stochastic block model is able to produce a wide variety of different network structures, including traditional assortative communities and different from them disassortative structures. In theory, it should be possible to guide the inference algorithm which type of structures is preferred when both types of structures fit the stochastic block model and its degree-corrected variant well for the input network. However, the existence of multiple local optima of the log-likelihood traps the inference algorithms in one of them. Although there were efforts to enable community detection at different levels of granularity, cf.~\cite{fortunato2018multi,lu2019asymptotic}, the need for controlling assertiveness of the solution has not been addressed so far.
Here, we apply a simple yet effective constraint on nodes' internal degree ratio in the objective function. This approach is independent of the inference algorithm. The resulting algorithm reliably finds assortative or disassortive structure as directed by the value of a single parameter $f$. We validated the model experimentally testing its performance on several real and synthetic networks. The experiments show that the inference of degree-corrected stochastic block model quickly converges to the sought-after assortative or disassortative structure. In contrast, the inference of degree-corrected stochastic block model gets often trapped in the inferior local optimal partitions.

\section*{Methods}
\subsection*{Proving properties of the Regularized Stochastic Block Model}

\begin{theorem} \label{thm:1}
When $f_i = 1/2$ for each node $i$ in the network, the MLE of our RSBM model defined by Eq.~\ref{eq:important2} becomes the MLE of degree-corrected stochastic block model\footnote{The proofs of Theorem~\ref{thm:1},\ref{thm:2} and \ref{thm:3} are provided in the Supplementary Material.}.
\end{theorem}

\begin{figure*}[!htb]
    \centering
    \includegraphics{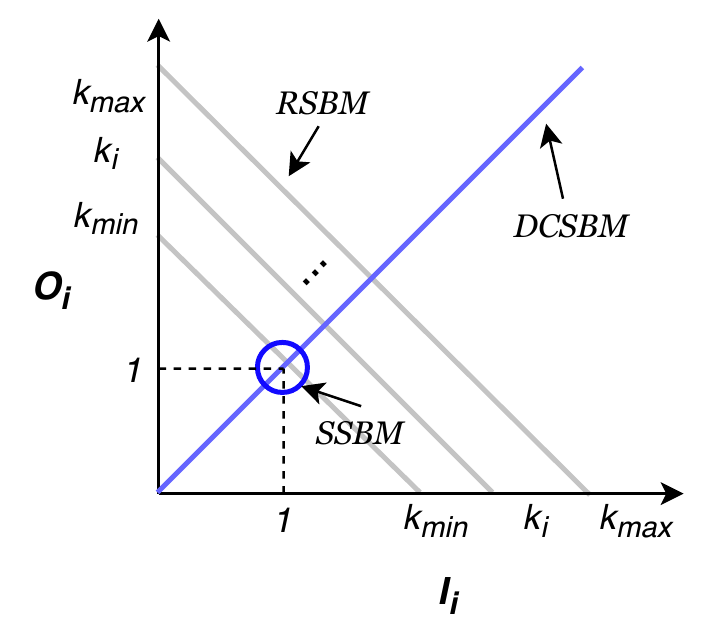}
    \caption{A unified view on the variants of stochastic block models. The standard stochastic block model (SSBM) is represented by the dark blue point $(1, 1)$ since it sets $I_i=O_i=1$ for each node $i$. The degree-corrected stochastic block model (DCSBM) requires $I_i = O_i$, which corresponds to the light blue line with a slope $f_i=1/2$ in the 2D coordinate plane. The introduced here regularized stochastic block model (RSBM) allows choosing $I_i$ and $O_i$ on the line $I_i + O_i = k_i$ where $k_i$ is the node degree. So our model is represented by a set of points on the parallel gray lines each of which intersects the x and y-axis at integer values corresponding to the degree of a node in the range $[k_{\text{min}}, k_{\text{max}}]$.}
    \label{fig:sbm_type}
\end{figure*}

According to Theorem~\ref{thm:1}, if we add a constraint $I_i = O_i$ for every $i$, the model introduced here reduces to the degree-corrected stochastic block. Figure~\ref{fig:sbm_type} illustrates the relationship between the most commonly used variants of stochastic block models. If we plot the constraints of these models on a 2D plane with $I_i$ and $O_i$ as the $x$ and $y$ axes, respectively, the standard stochastic block model (SSBM) sets $I_i=O_i=1$ for each $i$, which maps to the point (1,1) in the 2D coordinate plane. The constraint $I_i = O_i$ which represents the degree-corrected stochastic block model (DCSBM) maps onto the line with a slope $f_i=1/2$ in the 2D coordinate plane. Here, we extend the stochastic block model in two different directions: (i) using a constraint on $\theta_i = I_i + O_i$; (ii) choosing $I_i$ and $O_i$ on the line with a customized slope $f_i$. It turns out the second choice, like degree-corrected
stochastic block model, preserves the degree sequence of the network.


\begin{theorem} \label{thm:2}
Given any customized $\{f_i\}$ and the corresponding maximum-likelihood estimator $\hat{\theta}_i$, our model defined by Eq.~\ref{eq:important2} preserves the degree sequence of a network, so the expected degree of node $i$ generated by the RSBM model with $\hat{\theta}_i$ is
\begin{equation}
    \sum_j \lambda_{ij} = k_i
\end{equation}
\end{theorem} 

\subsection*{Information-theoretic interpretations}

While the maximum likelihood estimator $\hat{\theta}_i$ preserves the degree sequence in the observed network, we find that the closed-form analytic expression of such estimator is hard to obtain. However, when we impose constraints that for each $i$, $\theta_i = k_i$ and look for the MLE of $f_i$ for every $i$, the log-likelihood of the model introduced here has an interesting interpretation uncovered by the following theorem.

\begin{theorem} \label{thm:3}
When $\theta_i = k_i$ for every node $i$, maximizing the log-likelihood of Eq.~\ref{eq:important2} is equivalent to maximizing the target function
\begin{align} \label{eq:meaning0}
    \mathcal{L} = \mathbb{D}_{KL}(p_{\text{degree}}(r,s) || p_{\text{null}}(r,s)) - 2 \mathbb{E}_{k_i}[ H(\frac{k_i^+}{k_i}) ]
\end{align}
where the first term represents the Kullback–Leibler divergence between the edge distribution under the block model and the corresponding distribution under the null model which randomizes the edges inside and across blocks respectively, while the second term is twice the expectation of the binary entropy of in-degree ratio $k_i^+/k_i$.
\end{theorem}

Imposing the constraint $I_i + O_i = k_i$, we obtain a new model which involves a regularization term of the expected entropy of in-degree ratio $k_i^+/k_i$. According to Theorem~\ref{thm:3}, maximizing the log-likelihood of this model has a physical interpretation: the MLE of the RSBM model increases the KL divergence between the edge distribution under the block model and the corresponding edge distribution under the \textit{null} model, and, at the same time, it decreases the expected entropy of the in-degree ratio $k_i^+/k_i$. Intuitively, the \textit{null} model here separates the edges inside a block and the edges across different blocks. It assumes all the edges inside blocks are statistically equivalent, and so are all the edges across different blocks. In contrast, the \textit{null} model of the degree-corrected stochastic block model~\cite{karrer2011stochastic} mixes all the edges, assuming all the edges are statistically equivalent - the KL divergence term in its \textit{null} model does not distinguish the edges inside the same block and those across different blocks. On the other hand, the sum of entropy function $H(k_i^+/k_i)$ controls the identification of the edges' types. In Eq.~\ref{eq:meaning0}, $H(k_i^+/k_i)$ is the entropy function which achieves its minimum when $k_i^+/k_i \to 0$ or $k_i^+/k_i \to 1$. Therefore, the MLE of this model is likely to classify the edges of a node as either all in-edges or all out-edges -- the former case is likely to detect assortative communities and the latter case infers disassortative structure as the most probable block assignment. 

For community detection problems, it is rarely observed that all neighbors of a node are not located in the same block, especially for those nodes with low degrees. Unlike traditional community detection algorithms, the definition of stochastic block model and its variants mentioned above does not explicitly control the portion of neighbors in different blocks, rather they rely on the specific statistical inference to determine the block assignments. This observation inspires us to regularize the traditional stochastic block model on the in-degree ratio. Specifically, our \textbf{r}egularized \textbf{s}tochastic \textbf{b}lock \textbf{m}odel (\textbf{RSBM}) introduced here maximizes the objective function of Eq.~\ref{eq:important2} with $\theta_i = k_i$ and $f_i$ defined by the prior in-degree ratio. As explained above, the KL divergence of RSBM has similar meaning as that of degree-corrected stochastic block model. The expected cross-entropy term serves as a regularization term to control the resulting partition. Intuitively, when $f_i$ is close to one, then the inference algorithm tends to cluster nodes in a network into dense modules; when $f_i$ is close to zero, then disassortative partitions such as the core-periphery structure are likely to be discovered because the regularized block model tends to assign adjacent nodes into different blocks to decrease $k_i^+$.

\section*{Data availability statement}
The network data sets used for the evaluation of our proposed model are available at \url{http://konect.uni-koblenz.de/networks/}


\onecolumn
\section*{Acknowledgment}
The authors express thanks to Dr. Santo Fortunato for discussions of degree-correlated SBM. This work was supported in part by the Army Research Laboratory (ARL) through the Cooperative Agreement (NS CTA) Number W911NF-09-2-0053, and by the Office of Naval Research (ONR) under Grant N00014-15-1-2640. The views and conclusions contained in this document are those of the authors and should not be interpreted as representing the official policies either expressed or implied of the Army Research Laboratory or the U.S. Government.

\section*{Author contributions}
All authors contributed to the study design and manuscript preparation, XL conducted simulations and contributed the proof of the properties of the regularized block model introduced here, BKS contributed the functional forms of the prior in-degree ratios. Both XL and BKS reviewed and wrote the manuscript.

\section*{Competing interests}
Authors declare no conflict of interest as defined by Nature Research, or other financial and non-financial interests that might be perceived to influence the results and/or discussion reported here.

\section*{Supplementary Information} \label{sec:supplement}
\subsection*{Review of degree-corrected stochastic block model}
Since the standard stochastic block model makes nodes in the same block statistically indistinguishable, the most likely block assignment often places the nodes of similar degrees in a block, resulting in blocks of nodes with homogeneous degrees, too limited to represent the traditional community structures.

The degree-corrected stochastic block model~\cite{karrer2011stochastic} clusters together the nodes with heterogeneous degrees and defines the expected number of edges between nodes $i$ and $j$ as $\lambda_{ij} = \omega_{g_i,g_j} \beta_i \beta_j$ where for each node $l$, $g_l$ is its block assignment and $\beta_l$ is a model parameter associated with it. This simple yet effective extension improves the performance of the models for statistical inference of block structure in the real-world networks because nodes in a community tend to have broad degree distributions. To simplify technical derivations, authors in \cite{karrer2011stochastic} approximate the \textit{Bernoulli} distribution with the \textit{Poisson}. This approximation is tight when the product of the number of nodes $n$ and edge probability $p$ is small. Given a partition of the network, i.e. its block assignments $\{g_i\}$, reference~\cite{karrer2011stochastic} obtains the maximum likelihood estimates of the model parameters equal to $\hat{\beta}_i = \frac{k_i}{\kappa_{g_i}}$ and $\hat{\omega}_{rs} = m_{rs}$, where $k_i$ is the degree of node $i$, $\kappa_{r}$ is the sum of the degrees of all nodes in a block $r$, and $m_{rs}$ is the total number of edges between blocks $r$ and $s$, or, if $r=s$, twice the number of edges in $r$. Note that the self-loop edges and multi-edges are considered here, and the derivation is shown in Eq.~\ref{eq:ll0}.

Using the log-likelihood in Eq.~\ref{eq:ll0} and~\ref{eq:ll} leads to a new criterion to be optimized over all possible block assignments. Ignoring the constant terms which are irrelevant to the statistical inference of $\bf g$, the log-likelihood can be expressed as
\begin{equation} \label{eq:dcsbm_II}
    \mathcal{L}( G|{\bf g}) = \sum_{rs} m_{rs} \log \frac{m_{rs}}{\kappa_{r}\kappa_{s}} ,
\end{equation}
where $m_{rs}$ is the total number of edges between blocks $r$ and $s$ if $r \neq s$, or twice that number otherwise, and $\kappa_{r} = \sum_{i:g_i=r} k_i$ is the sum of degrees of the nodes in block $r$. This criterion has an information-theoretic interpretation when it is presented in the alternative form 
\begin{equation}
     \mathcal{L}(G|{\bf g}) = \sum_{rs} \frac{m_{rs}}{2m} \log \frac{m_{rs}/{2m} }{(\kappa_{r}/2m)(\kappa_{s}/{2m})} ,
\end{equation}
which is the Kullback–Leibler divergence between the probability of observing an edge $(i, j)$ between block $r$ and $s$ in real network, i.e. $P[i\in r,j \in s] = m_{rs}/{2m}$, and the probability of observing such edge in a random graph with the same degree sequence, i.e. $P' [i \in r,j \in s] = (\kappa_{r}/2m)(\kappa_{s}/{2m})$. Seen this way, the statistical inference maximizes the information gain when replacing the generative model of a random graph model with the degree-corrected stochastic block model. In other words, it requires the most information to describe the observed graph starting from a random graph that does not have block structure at all.

It is worth mentioning that, in the standard stochastic block model where the number of edges is drawn from the \textit{Bernoulli} distribution, it is rare that $\omega_{rs}$ is close to 1. This is because the numbers of the edges between blocks are usually very small when the network is sparse. A \textit{Bernoulli} random variable with a small mean is well approximated by a \textit{Poisson} random variable having the same mean~\cite{perry2012null}. For this reason, \cite{zhao2012consistency} found no significant difference in the solutions produced by this model using the \textit{Poisson} distribution compared to the original model using the \textit{Bernoulli} distribution. Hence here, we refer to the standard stochastic block model as the one in which the number of edges between two nodes is drawn from the \textit{Poisson} distribution.

\subsection*{Markov chain Monte Carlo} \label{sec:infer}
We use the Markov chain Monte Carlo (MCMC) algorithm~\cite{nasrabadi2007pattern} to infer the block assignment of the stochastic block model and its extensions. The MCMC algorithm samples the distribution of all possible block assignments in the network partitions space. Suppose there are $K$ blocks and $N$ nodes in a network. The naive Markov Chain Monte Carlo approach is not practical because the size of the partition space $O(N^K)$ which grows exponentially in the number of blocks $K$. Therefore, Piexoto ~\cite{peixoto2014efficient} proposes the optimized MCMC algorithm with the greedy heuristic to infer the block assignment. Initially, every node in the network is assigned to one random block independently. Then, one attempts to move a node from block $r$ to $s$ with a probability conditioned on its neighbor's block assignment $t$. The proposal function is defined as
\begin{equation} \label{eq:proposal} 
    p(r\to s|t) = \frac{m_{ts} + \epsilon}{\sum_s m_{ts} + \epsilon B}
\end{equation}
where $\epsilon > 0$ is a free parameter to fulfill the ergodicity condition such that any block assignment can be reached from the current block assignment in the finite and aperiodic expected number of steps. When $\epsilon \to \infty$, the proposal function reduces to the naive scheme which assign a random block to the current node. However, in such case, the possibility of current node being assigned to the correct block assignment is very low, thus, it does not increase the log-likelihood in most cases. Consequently, such proposals are rejected very frequently, wasting the computational resource. By applying a relatively small $\epsilon$, the proposal function defined above is more likely to get accepted. The intuition behind this proposal function is that, given that a large number of edges laying across blocks $s$ and $t$, a node with many neighbors in block $t$ is likely to reside in block $s$. Consequently, the proposal function defined by Eq.~\ref{eq:proposal} is more likely to be accepted, saving the computational cost for many rejected proposals.

To ensure the detailed balance, each proposed assignment is accepted with a probability $a$ in the Metropolis-Hastings fashion~\cite{metropolis1953equation} given by
\begin{equation} \label{eq:tiago_a}
a = \min \Bigl\{ \exp(\Delta \mathcal{L}) \frac{\sum_t n_t p(s\to r|t)}{\sum_t n_t p(r\to s|t)} \Bigr\}  ,    
\end{equation}
where the node of the proposed assignment has $n_t$ neighbors in block $t$ and $\Delta \mathcal{L}$ is the change of log-likelihood after each proposal. A list of edges which are adjacent to each block are used to compute Eq.~\ref{eq:tiago_a}, which can be done in $O(k_i)$ time~\cite{peixoto2014efficient} where $k_i$ is the degree of node $i$. Thus, an entire MCMC sweep of all nodes in the network requires $O(E)$ operations where $E$ is the number of edges in the networks. The mixing time of the MCMC algorithm depends on the specific input network and the starting point.

\subsection*{Proof of Theorem 1}
\begin{proof}
If $f_i = 1/2$ for each $i$, then $I_i = O_i = \theta_i / 2$ and $\sum_{i\in r} I_i = \sum_{i\in r} O_i = \sum_{i\in r} \theta_i / 2$. Hence, $\Lambda_{rs} = \Theta_r/2 \times \Theta_s/2$ where $\Theta_r = \sum_{i \in r} \theta_i$. Also, it is obvious that the sum of cross entropy on the RHS of Eq.~\ref{eq:important2} is a constant term given the block assignment $\bf g$. The log-likelihood of our model defined by Eq.~\ref{eq:important2} can be rewritten in the form
\begin{align*}
    \mathcal{L}(G|{\bf g,I,O})
    &= \sum_{rs} m_{rs} \log \frac{m_{rs}}{\Theta_r \times \Theta_s} + 2\sum_i k_i \log \theta_i + const.
\end{align*}
where all the constant terms are grouped into the last term on the RHS. Setting the derivative of the RHS over $\theta_i$ to zero, we obtain $k_i/\theta_i = \sum_s m_{g_i, s} / \Theta_{g_i}$ which means that for every node $i$ the MLE estimator $\hat{\theta}_i = \Theta_{g_i} k_i / \kappa_{g_i}$. Plugging $\hat{\theta}_i$ into the log-likelihood above, we obtain the log-likelihood expression in Eq.~\ref{eq:dcsbm_II} after dropping all the constant terms.
\end{proof}

\subsection*{Proof of Theorem 2}
\begin{proof}
For given $\{f_i\}$, the derivative of the log-likelihood defined in Eq.~\ref{eq:important} over $\theta_i$ is zero when the maximum likelihood estimator $\hat{\theta}_i$ satisfies the condition
\begin{equation} \label{eq:preserve_deg}
    \frac{k_i}{\hat{\theta}_i} = \frac{m_{rr}}{\sum_{i\in V_r} f_i \hat{\theta}_i} f_i + \frac{\sum_{s:s\neq r} m_{rs}}{\sum_{i\in V_r} (1-f_i) \hat{\theta}_i} (1-f_i)
\end{equation}
Even without the closed-form expression of $\hat{\theta}_i$, we show that Eq.~\ref{eq:preserve_deg} ensures that the expected degree of node $i$ is
\begin{align} 
    \sum_{j} \lambda_{ij} &=  I_i \sum_{j \in g_i}I_j \frac{m_{g_i,g_i}}{\lambda_{g_i,g_i}} + O_i \sum_{j \notin g_i} O_j \frac{m_{g_i,g_j}}{\lambda_{g_i,g_j}} \\
    &= I_i \frac{m_{g_i,g_i}}{\sum_{j\in g_i} I_j} + O_i \frac{ \sum_{b:b\neq g_i} m_{g_i,b}}{\sum_{j\in g_i} O_j}  \label{eq:expected_degree}
\end{align}
Since $\theta_i = \hat{\theta}_i$, we have $I_i = f_i \hat{\theta}_i$ and $O_i = (1 - f_i) \hat{\theta}_i$, so the RHS of Eq.~\ref{eq:expected_degree} is actually the RHS of Eq.~\ref{eq:preserve_deg} multiplied by $\hat{\theta}_i$, hence it is equal to observed node degree $k_i$ in the network. Therefore, $\sum_{j} \lambda_{ij} = k_i$.
\end{proof}

\subsection*{Proof of Theorem 3}
\begin{proof}
We start with the equivalent definition of the generalized model in Eq.~\ref{eq:important}. The first-order derivatives of the log-likelihood over $I_i$ and $O_i$ are
\begin{align}
\frac{\partial \mathcal{L}}{\partial I_i} &= - m_{g_i g_i} \frac{2 \sum_{i\in g_i} I_i }{(\sum_{i\in g_i} I_i)^2} + \frac{2 k_i^+}{I_i} \\
\frac{\partial \mathcal{L}}{\partial O_i} &= - \sum_{s:s\neq g_i} m_{g_i s} \frac{2 \sum_{i\in s} O_i }{\sum_{i\in g_i} O_i \cdot \sum_{i\in s} O_i} + \frac{2 k_i^-}{O_i}  
\end{align}
Given the block assignment $\bf g$, when log-likelihood achieves its maximum, the stable points of $I_i$ and $O_i$ satisfy $\frac{\partial \mathcal{L}}{\partial I_i} = \frac{\partial \mathcal{L}}{\partial O_i} = 0 $, which leads to the equations
\begin{equation}
    \forall i \in r: \quad\quad \frac{k_i^+}{I_i} \sum_{i \in r} I_i + \frac{k_i^-}{O_i} \sum_{i \in r} O_i = \sum_{i \in r} I_i + O_i = \kappa_r
\end{equation}
Suppose there are $R$ nodes in block $r$, so we have $R$ independent equations shown above and the other $R$ equations $I_i + O_i = \theta_i = k_i$ representing the assumed constraints. These $2R$ independent equations produce at most one unique set of solutions for the $2R$ variables. It is obvious that $I_i = k_i^+$ and $O_i = k_i^-$ satisfy all these $2R$ equations. Therefore, Eq.~\ref{eq:important2} can be rewritten as
\begin{equation}
    \mathcal{L} = \sum_{rs} \frac{m_{rs}}{2m} \log \frac{m_{rs} / 2m}{\Lambda_{rs} / (2m)^2} - 2 \sum_i \frac{k_i}{2m} (H(\frac{k_i^+}{k_i}))
\end{equation}
where $\Lambda_{rs} = \kappa_r^+ \kappa_r^+$ if $r=s$;  $\Lambda_{rs} = \kappa_r^- \kappa_s^-$ otherwise. Here $\kappa_{r}^{+/ -} = \sum_{i\in r} {k_i^{+/ -}}$ are the sum of internal and external degrees of nodes respectively.

The first sum on the RHS of above equation represents the Kullback–Leibler divergence between the distribution of edges residing across blocks $r$ and $s$, i.e., $p_{\text{degree}}(g_i= r,g_j = s) = \frac{m_{rs}}{2m}$ and the corresponding \textit{null} distribution is defined as 
\begin{align}
    p_{\text{null}}(g_i= r,g_j= s) = \begin{cases} &\frac{\kappa_{r}^-}{2m} \frac{\kappa_{s}^-}{2m} \quad \text{if} \quad r \neq s\\
    & \frac{\kappa_{r}^+}{2m} \frac{\kappa_{s}^+}{2m} \quad \text{if} \quad r = s
    \end{cases}
\end{align}
Note that the \textit{null} model here does not require that all the edges across the entire network are randomly distributed. But it does require that, the edges inside each block and across different blocks are so distributed.

Since $\sum_i k_i = 2m$, the term $\frac{k_i}{2m}$ can be treated as the probability of selecting an edge of node $i$. Thus, the second sum on the RHS can be treated as the expectation of binary entropy function $H(\frac{k_i^+}{k_i})$, i.e., $\mathbb{E}_{k_i}[ H(\frac{k_i^+}{k_i}) ] = \sum_i \frac{k_i}{2m} (H(\frac{k_i^+}{k_i}))$.
\end{proof}

\subsection*{Real networks for experiments}
Table~\ref{tab:realnetworks} shows the number of nodes and edges of the real networks used for experiments. The number of blocks is set to their values generally accepted in the previous publications. As shown in Table~\ref{tab:realnetworks}, the Karate club network, Dolphin social network and Les Miserables characters network are relatively sparse because their average degrees range from 4 to 6. To investigate the performance of the regularized model, we add ``noises'' to these graph by randomly removing their edges, further increasing the sparsity of these networks. For each network, we iteratively remove random edges from the original set of edges and repeat this process twice (each time starting from the same original network). If the removal of an edge results in the isolated nodes in the network, we remove it  along with its edges.

\begin{table*}[!htb]
    \centering
    \caption{Real networks for experiments. The number of blocks for each network is set to its value generally accepted in the previous publications.}
    \label{tab:realnetworks}
    \begin{tabular}{ccccc}
    \hline
    Network & \#Nodes & \#Edges & \#Blocks & Ref.\\
    \hline
    Karate club & 34 & 78 & 2 & \cite{zachary1977information}\\
    Dolphin social network & 62 & 159 & 2 & ~\cite{lusseau2003bottlenose}\\
    Characters from Les Miserables & 77 & 254 & 6 & ~\cite{newman2004finding} \\
    \hline
    \end{tabular}
\end{table*}

\subsection*{Dimensionality reduction of the partition space}
In Figures~\ref{fig:dcsbm_3dscatter} and~\ref{fig:gsbm_3dscatter}, we project the multidimensional space of partitions into the 2D x-y plane. Specifically, for each MCMC trial, we record all the sampled partitions of the network and their corresponding log-likelihoods. Given two sampled partitions $\{g_i\}$ and $\{b_i\}$, a new partition $\{l_i\}$ is randomly generated such that $l_i = g_i$ with probability $0.5$ or $l_i = b_i$ otherwise. Then, we scatter all the sampled partitions along with the randomly generated partitions on the x-y plane of Figure~\ref{fig:dcsbm_3dscatter}. In this figure, the multiple-dimensional partition space is projected to a two-dimensional x-y plane by the multidimensional scaling (MDS) algorithm~\cite{borg2017applied}. The z-axis indicates the log-likelihood of the partitions.

\subsection*{Illustrative toy networks}
\begin{figure*}[!htb]
\centering
\subfloat[Core-periphery]{ \makebox[5cm][c]{ \includegraphics[width=3cm]{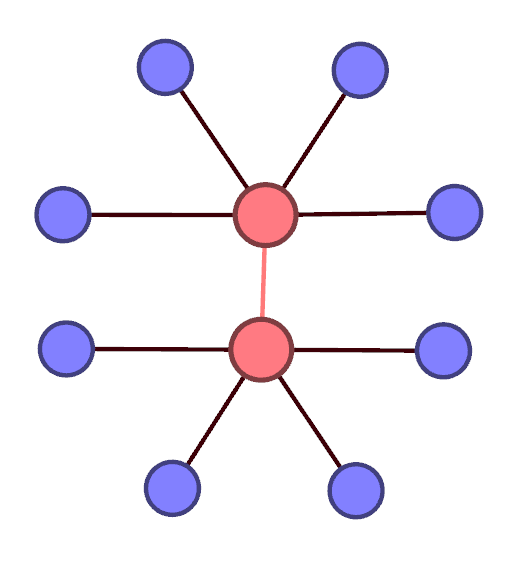}} }
\subfloat[Twisted Core-periphery]{\makebox[5cm][c]{ \includegraphics[width=3cm]{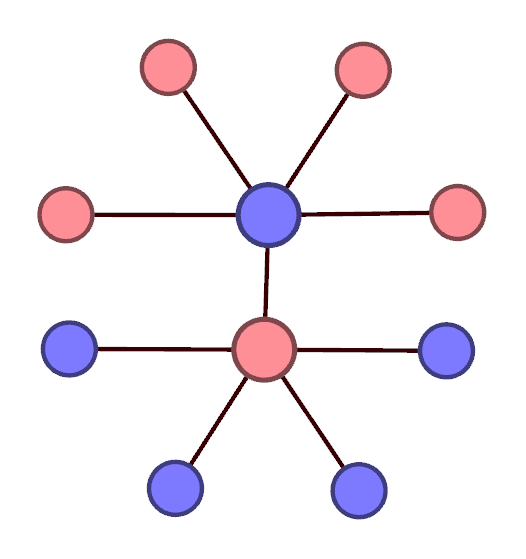}} }
\subfloat[Assortative Community]{\makebox[5cm][c]{ \includegraphics[width=3cm]{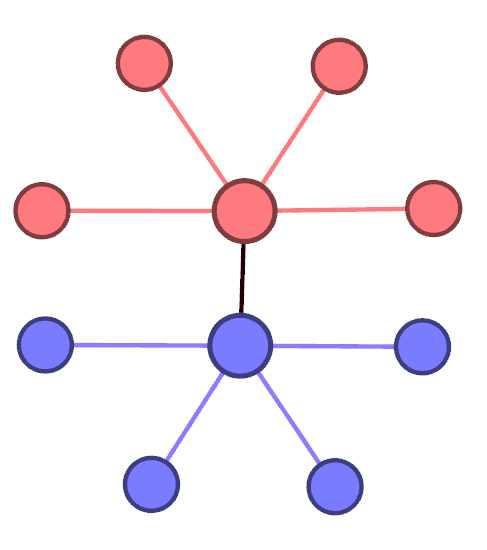}} }
\caption{The twin stars network and three possible partitions. The color represents a node's block assignment. \textbf{(a)} Core-periphery partition with coverage $0.11$; \textbf{(b)} Twisted core-periphery partition with coverage $0$; \textbf{(c)} Assortative community partition with coverage $0.89$.}
\label{fig:twin_star}
\end{figure*}

We start with an illustrative toy network to demonstrate the effect of regularization introduced here on the stochastic block model. Consider the ``twin stars" network as shown in Figure~\ref{fig:twin_star}. 

\begin{table*}[!htb]
    \centering
    \caption{The log-likelihoods of the three partitions of the twin stars network under the standard stochastic block model (SSBM), the degree-corrected stochastic block model (DCSBM) and our block model with regularization (RSBM). The partitions are shown in Figure~\ref{fig:twin_star} and the corresponding maximum log-likelihood values are marked by the bold font.}
    \label{tab:twin_star_ll}
    \begin{tabular}{cccc}
         \noalign{\hrule height 0.7pt}
         Model & Core-periphery & Twisted core-periphery & Assortative Community\\
         \hline
         SSBM & \textbf{-12.47664} & -18.38972 & -24.66870 \\
         DCSBM & -44.66540 & \textbf{-39.55004} & -45.82902 \\
         RSBM ($\alpha=0.3$) & -20385.51476 & -377.52624 & \textbf{-44.68795} \\
         RSBM ($\alpha=0.6$) & -20394.46861 & -368.57239 & \textbf{-43.38465} \\
         RSBM ($\alpha=0.9$) & -20416.64932 & -346.39168 & \textbf{-43.35718} \\
         \noalign{\hrule height 0.7pt}
    \end{tabular}
\end{table*}

There are three possible partitions of this twin stars network. The first option is to cluster the two central hubs into one block and the remaining nodes into another block. This structure corresponds to a core-periphery structure as shown in Figure~\ref{fig:twin_star}(a) where the color of a node represents its block assignment. If we exchange the block assignments of the nodes connected to the two hubs, we obtain a twisted core-periphery structure as shown in Figure~\ref{fig:twin_star}(b). Both of these two partitions are disassortative. For community detection problems, however, the most suitable partition should divide the network into two identical stars, which corresponds to the partition shown in Figure~\ref{fig:twin_star}(c).

Table~\ref{tab:twin_star_ll} lists the log-likelihoods of these three partitions under the standard stochastic block model (SSBM), the degree-corrected stochastic block model (DCSBM) and our block model with regularization terms (RSBM). Under the degree-corrected stochastic block model, the log-likelihood of the assortative community partition is smaller than the log-likelihood of the inappropriate twisted core-periphery partition, whereas our new model manages to identify the assortative community partition as the most appropriate one. We tested the log-likelihood of the regularized stochastic block model with the $f(k)$ function using the model parameter $\alpha=0.3$, $0.6$ and $0.9$, respectively. Note that, in Eq.~\ref{eq:important2}, $\Lambda_{rs} = 0$ if a block $r$ includes only degree-one nodes, causing a log-likelihood of negative infinity according to the model definition. Thus, we set the term $m_{rs} \log m_{rs}/\Lambda_{rs} = -10,000$ in Eq.~\ref{eq:important2} to ensure the log-likelihood is small enough. In general, larger $\alpha$ values would result in a larger gap in the log-likelihoods of the assortative community partition and the other two partitions. However, all the possible $\alpha$ values ensure that the assortative community partition fits the regularized model introduced here much better than the other candidates. The reason is that, no matter what value of $\alpha$ gets selected here, we have $f(1) = 1$. To reduce the cross entropy terms in Eq.~\ref{eq:meaning0}, the inference algorithm is likely to assign a degree-one node to the block in which its sole neighbor resides. Therefore, the inference algorithm is likely to return the assortative communities as the most likely partition of the network.

\end{document}